\documentclass{kluwer}
\usepackage{graphicx} 

\DeclareMathAlphabet{\mathbfl}{OML}{cmm}{b}{it} 

\newcommand{\DS}{\ensuremath{\displaystyle}}

\newcommand{\abs}[1]{\ensuremath{| #1 |}}

\newcommand{\mathtext}[1]{\ensuremath{\;\mbox{#1}\;}}

\newcommand{\Derv}[2]{\ensuremath{\frac{d #1}{d #2}}}

\newcommand{\Dpar}[2]{\ensuremath{\frac{\partial #1}{\partial #2}}}

\begin{document}
\begin{article}
\begin{opening}
\title{Magnetic modelling and tomography:\\
 First steps towards a consistent
reconstruction of the solar corona}

\author{T. \surname{Wiegelmann}\email{wiegelmann@linmpi.mpg.de}}
\author{B. \surname{Inhester}} 
\institute{Max-Planck-Institut f\"ur Aeronomie,
Max-Planck-Strasse 2, 37191 Katlenburg-Lindau, Germany}
\date{DOI: 10.1023/A:1024282131117 \\
Bibliographic Code: 2003SoPh..214..287W}



\runningtitle{Magnetic modelling and tomography}
\runningauthor{Wiegelmann and Inhester}

\begin{ao}
Kluwer Prepress Department\\
P.O. Box 990\\
3300 AZ Dordrecht\\
The Netherlands
\end{ao}

\begin{motto}

\end{motto}

\begin{abstract}
We undertake a first attempt towards a consistent reconstruction of the
coronal magnetic field and the coronal density structure.
We consider a stationary solar corona which has to obey the equations of
magnetohydrostatics.
We solve these equations with help of a newly developed
optimization scheme.
As a first step we illustrate how tomographic information can be included
into the reconstruction of coronal magnetic fields. In a second step we use
coronal magnetic field information to improve the tomographic inversion
process. As input the scheme requires magnetic field measurements on
the photosphere from vector-magnetographs and the line-of-sight integrated
density distribution from coronagraphs.
We test our codes with well known analytic magnetohydrostatic
equilibria and models.
The program is planed for use within the STEREO mission.
\end{abstract}


\abbreviations{\abbrev{KAP}{Kluwer Academic Publishers};
   \abbrev{compuscript}{Electronically submitted article}}

\nomenclature{\nomen{KAP}{Kluwer Academic Publishers};
   \nomen{compuscript}{Electronically submitted article}}

\classification{JEL codes}{D24, L60, 047}
\end{opening}
\section{Introduction}
The solar magnetic field is an important quantity which couples the solar
interior, the photosphere and the atmosphere. The quasi stationary coronal
magnetic field configuration is an interesting and challenging topic on its
own right. But even to
understand basic processes like coronal mass ejections and flares it is
important to understand the quiescent magnetic configuration out of which
these dynamic phenomena arise.
Unfortunately the coronal magnetic field cannot be measured directly, but it
has to be reconstructed from photospheric measurements.
 A magnetic field reconstruction of the solar corona
has to be consistent with the observed spatial variation of the coronal plasma (density,
pressure, temperature) often elongated along the magnetic field.

Here we are mainly interested in long living
structures which are time independent in first order. We also concentrate on
closed magnetic configuration where a stationary plasma flow (solarwind) does
not significantly contribute to the force balance. Such configurations
 are static equilibria and have to obey the magnetohydrostatic
equations (MHS).

As the magnetic field ${\bf B}$ and the density
distribution $N$ are physically closely related their model reconstruction
should also be linked as much as possible. In this paper we attempt to
show how this can be achieved. We propose variational principles which
if they can be solved should give a consistent model for an isothermal corona.
For the magnetic field reconstruction this
leads to a generalization of a nonlinear force-free approach by
 \cite{wheatland00}, for the density reconstruction
we obtain a tomography problem with an improved regularization term.

The ground based or space-born magnetograph observation provide either the line-of-sight
magnetic field ($B_{los}$, e.g. MDI on SOHO), which is sufficient for potential and
linear force-free fields, or all three components of the photospheric
magnetic field (e.g. IVM in Hawaii, expected also from SolarB). The latter information
is sufficient to determine nonlinear force-free fields completely. As a
force-free approximation is justified only in the limit of a vanishing plasma $\beta$, we
take into account forces (pressure gradient and gravity) for configurations with a
finite plasma $\beta$ even though we shall consider $\beta$ small.

Popular simplifications for the reconstruction of coronal magnetic fields
are:
\begin{itemize}
\item Potential fields (${\bf j}= {\bf 0}$)
(e.g.\ \opencite{schmidt64}; \opencite{semel67}; \opencite{schatten69};
\opencite{sakurai82}; \opencite{rudenko01a})
\item Linear force-free fields
(e.g.\
\opencite{nakagawa72}; \opencite{chiu77};
\opencite{seehafer78}; \opencite{semel88}; \opencite{gary89};
\opencite{lothian95})
\item Linear non force-free fields
(e.g. \opencite{zhao93}; \opencite{zhao94};
\opencite{petrie00}; \opencite{zhao00}; \opencite{rudenko01b})
\item Nonlinear force-free fields
 (e.g.\
\opencite{sakurai81}; \opencite{wu85}; \opencite{roumeliotis96};
\opencite{amari97}; \opencite{mcclymont97}; \opencite{wheatland00};
\opencite{yan00}).
\end{itemize}
Within this work we do not use any of these assumptions but consider the
general case of nonlinear non-force-free equilibria. The mathematical problem
of calculating nonlinear non-force-free fields is closely related to the
problem of calculating nonlinear force-free fields which coincides with the
above in the limit of $\beta \rightarrow 0$. Under ideal conditions the
information contained in a (perfect) vectormagnetogram together with the force-free
condition would be sufficient to calculate the coronal magnetic field. Within
this work we  show that the
information contained in a vectormagnetogram together with a
tomographic reconstructed coronal density distribution and the assumption
of magnetohydrostatic force balance is as well sufficient
to calculate the  finite $\beta$ coronal magnetic field. Unfortunately current
vectormagnetograms and tomographic reconstruction are far from being
perfect, which affects the quality of reconstruction. Within this work we
use well known MHS-equilibria to test our newly developed reconstruction
program. The use of analytic equilibria as artificial data allows us to
extract ideal vectormagnetograms as well as ideal coronal density
distributions.

As for the density observations, ground based coronagraphs (e.g., the Mark III
coronagraph on Hawaii, LASCO coronagraph on SOHO and the future STEREO
mission) provide the line of sight integrated
density structure of the solar corona from different relative viewpoints
as the Sun rotates. These measurements have been used for a 3D-reconstruction
of the coronal plasma distribution with help of tomographic methods
\cite{davila94,zidowitz99,frazin00,frazin02}. The major problems here
are:
\begin{itemize}
\item the assumption of stationarity of coronal structures as the Sun rotates,
\item the lack of data due to the occulted center of the image,
\item the nonideal viewing geometry caused by a slight tilt of the Sun's axis
with respect to the ecliptic.
\end{itemize}
These shortcomings generally enhance the intrinsic ill-posedness of the
tomography problem. The general approach to stabilize the reconstruction is
to smooth the solution by regularization. The prize to pay is a reduced
spatial resolution of the model depending on the quality of the data
and inconsistencies and ill-conditioning due to the above effects.
So far only very general, isotropic
regularization operators have been applied to coronal density reconstruction
problems.
Our approach to the density reconstruction in connection with the
reconstruction of the coronal magnetic field leads to a new regularization
operator which, as we demonstrate by test calculations, could yield
a better spatial resolution than conventional reconstructions.

The paper is outlined as follows. In section \ref{basic}  we describe the
basic equations and the newly developed algorithm of the magnetic field
reconstruction program in the case where the plasma density distribution $N$
is given.
Section \ref{tests} contains several test-runs where we
apply our code for the reconstruction of analytic MHS-equilibria.
In section \ref{tomoopt} we propose an algorithm for an improved
reconstruction of $N$ if some information of ${\bf B}$ is given.
This approach is tested and compared with conventional methods by
with the help of a two dimensional analytic coronal density distribution.
In the final section \ref{conclusions} we discuss how both methods could be
used together to derive a consistent model of the Sun's corona.
In appendix \ref{appendixA}, \ref{appendixB} and \ref{appendixC}
we provide the algebra which has been omitted in the text.

\section{Basic equations}
\label{basic}
We describe the coronal plasma with help of the magnetohydro static (MHS)
equations.
The MHS  equations are
\begin{eqnarray}
{\bf j}\times{\bf B} -\nabla P -\rho \nabla \Psi  & = & {\bf 0},
\label{forcebal}\\
\nabla \times {\bf B }& = & \mu_0 {\bf j}  \label{ampere}, \\
\nabla\cdot{\bf B}    & = &         0      \label{solenoidal},
\end{eqnarray}
where ${\bf B}$ is the magnetic field, ${\bf j}$ the electric current
density, $P$ the plasma pressure, $\rho$ the plasma density,
$\mu_0$ the vacuum permeability and $\Psi$ the solar gravity potential.
We define the functional
\begin{equation}
L=\int_{V} \left[B^{-2} \, |(\nabla \times {\bf B}) \times {\bf B} -\mu_0
(\nabla P +\rho \nabla \Psi)|^2 +|\nabla \cdot {\bf B}|^2\right] \; d^3x.
\label{defL}
\end{equation}

The domain $V$ is a volume which on one side is bounded by the sun's
photosphere.  Obviously, $L$ is bound from below by $0$.  This bound is attained
if the magnetic field satisfies the MHS equations. Here we assume that the
plasma pressure and the density are given. It is assumed that the corresponding
information will be provided by tomographic reconstruction of the solar corona.
We vary functional $L$ with respect to an iteration parameter $t$ and get (see
Appendix \ref{appendixA} for the derivation)
\begin{equation}
\frac{1}{2} \; \frac{d L}{d t}
=-\int_{V} \frac{\partial {\bf B}}{\partial t}\cdot {\bf F} \; d^3x
 -\int_{S} \frac{\partial {\bf B}}{\partial t} \cdot {\bf G} \; d^2x,
\label{minimize1} \end{equation}
where
\begin{eqnarray} {\bf F} & =&
\nabla \times ({\bf \Omega_a} \times {\bf B} )
- \bf \Omega_a \times (\nabla \times \bf B)
\nonumber\\
& & +\nabla(\bf \Omega_b \cdot \bf B)-  \bf
\Omega_b(\nabla \cdot \bf B) +( \Omega_a^2 + \Omega_b^2)\; \bf B,
\end{eqnarray}
\begin{eqnarray} {\bf G} & = & {\bf \hat n}
\times ({\bf \Omega_a} \times {\bf B} )
-{\bf \hat n} (\bf \Omega_b \cdot \bf B),
\end{eqnarray}
\begin{eqnarray}
{\bf \Omega_a} &=& B^{-2} \;\left[(\nabla \times {\bf B})
\times {\bf B} - \mu_0 (\nabla P +\rho \nabla \Psi) \right],
\nonumber\\
{\bf \Omega_b} &=& B^{-2} \;\left[(\nabla \cdot {\bf B}) \; {\bf B} \right].
\label{defomega}\end{eqnarray} %
The surface integral in (\ref{minimize1}) vanishes if ${\bf B}$
is prescribed on the boundary. We iterate the magnetic field inside the
computational box with
\begin{equation} \frac{\partial {\bf B}}{\partial t} =\mu
{\bf F}
\label{iterateB}, \end{equation}
which ensures that $L$ is monotonously
decreasing.  For the bottom the boundary values are given by the photospheric
vector magnetograph observations. On other boundaries we may either assume ${\bf
B}$ or include the boundary values in the variation.  Actually the handling of
the not observed lateral and top boundaries of the computational box is similar
here as in the nonlinear force-free case \cite{twtn03}.  On the boundary  of
the computational box the magnetic field is iterated with
\begin{eqnarray}
\frac{\partial {\bf B}}{\partial t} = & 0 \; &
\mbox{where ${\bf B}$ observed},
\\
\frac{\partial {\bf B}}{\partial t} = & \mu \; {\bf G} \; & \mbox{else}.
\label{eq11}
\end{eqnarray}

We propose to use this
iteration process to solve for the minimum of $L$. If a solution of the
MHS-equations for the prescribed boundary condition exist, the global
minimum of $L$ corresponds to this solution and attains $L=0$. Please note that the iteration
procedure ensures to find this global minimum if the solution space is
convex. For a non convex solution space it is possible that the iteration
will lead to a local minimum. For complicated magnetic field configurations
it is difficult to decide in advance whether the solution space is convex or
not. For a non convex  solution space it is still possible to find the global
minimum by iteration if the start configuration is sufficient close (within a convex area)
to this minimum. The method generalizes an approach by \cite{wheatland00}
which has been used to compute force-free fields.
\section{Convergence Tests}
\label{tests}
Since analytic truly 3D MHS equilibria are not available, we use
an analytic 2D MHS-equilibrium to test the newly developed code.
The analytic equilibria are not meant to be a good representation of
the solar corona and the tests are only carried out to check the convergence
of the newly developed code.
We represent the magnetic field with help of the flux-function $A(x,z)$ as
\begin{equation}
{\bf B}=\nabla A \times {\bf e_y} + B_y \; {\bf e_y}
\label{defB}
\end{equation}
and the MHS equations reduce to a Grad-Shafranov equation
\begin{equation}
- \Delta A=\frac{\partial}{\partial A} \left(P(A,\Psi) + \frac{B_y^2(A)}{2}
\right).
\label{gse}
\end{equation}
\subsection{Equilibrium MHS-1}
\begin{figure}
\includegraphics[width=12cm,height=10cm]{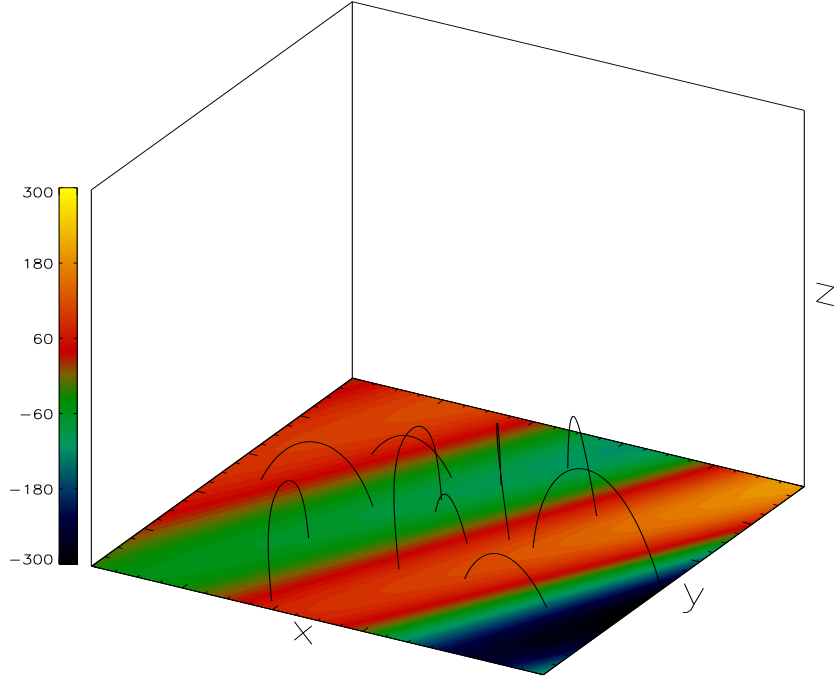}
\caption{MHS-1: Some field lines for the first equilibrium. The colours on the photosphere
correspond to the normal component of the magnetic field.}
\label{fig1}
\end{figure}
As a first test we consider equilibria without gravity $(P=P(A))$ and choose
$P(A) \propto A^2$ and $B_y(A) \propto A$. The corresponding Grad-Shafranov
equation is linear in $A$ and can be solved analytically  by a separation ansatz.
It is convenient to define a function
\begin{eqnarray}
\Pi(A) &=& c^2 \; A^2=P(A)+\frac{B_y^2(A)}{2}, \\
P(A) &=& a_0 \; c^2 \; A^2 \label{p1}, \\
B_y(A) &=&\sqrt{(1-a_0)} \; c \; A \label{by1}.
\end{eqnarray}
Configurations with $c=0$ correspond to potential
fields, finite $c$ and $a_0=0$ to force-free equilibria,
$a_0=1$ to equilibria with pressure gradient but without
magnetic shear and finally $0<a_0<1$ is the general case including both
forces and magnetic shear.
With this approach we get the solution of (\ref{gse}):
\begin{equation}
A(x,z) = \sum_{k=1}^{\infty} \exp(-\nu \pi z/L) \;
\left[a_k \cos(k \pi x/L) + b_k \sin(k \pi x/L) \right]
\label{A1},
\end{equation}
where $\nu=\sqrt{k^2-c^2}, \; \mbox{for } c^2<k^2$.
The solutions of the Grad-Shafranov equation are invariant in one spatial
coordinate ($\frac{\partial}{\partial y}=0$). To test our 3D-optimization code it
would be more convenient to have equilibria varying in all three spatial directions.
We construct such equilibria by rotating the solution of Grad Shafranov equation
by an angle $\phi_1$ around the
z-axis and by $\phi_2$ around the y-axis. As a result the solution varies in
all of our three coordinate directions.
The final equilibrium
has the following free parameters:
$c, a_0, a_k, b_k, \phi_1, \phi_2$.
As an example we choose
($c=0.8, a_0=0.5, a_1=1.0, a_3=-0.8,a_k=b_k=0$ for all other
$k$, $\phi_1=-0.05 \pi, \phi_2= 0.15 \pi$.)
With help of the flux function (\ref{A1}), the equations for the pressure
(\ref{p1}), shear field (\ref{by1}) and the magnetic field definition (\ref{defB}) we get
the magnetic field {\bf B} and the plasma pressure $P$. For an isothermal
plasma we derive the density as $\rho=\frac{P}{RT}$. We normalize the maximum
normal magnetic field at $z$ = 0 to 300 Gauss = 0.03 T.

As a first test, we want to reconstruct this equilibrium with our code.
The code needs any 3D-vector field as start configuration for the iteration
procedure and it is convenient to
choose a potential magnetic field with respect to the photospheric line
of sight magnetic field. The potential field can be easily computed for an
observed magnetogram. For our model data a potential field is computed
with the same parameter set but $c=0$.
The boundary values for the iteration are in practical cases only known
on the bottom plane. On the others plans they have to be iterated too using
(\ref{eq11}). In this test we fix  the magnetic field however everywhere on the
boundary to the value of the analytic solution to simplify the problem.
For the force-free case we treated the side and top boundaries
as unknowns in \cite{twtn03} and showed how they can be iteratively determined by (\ref{eq11}).
This latter way of treating the boundary values makes the finding of a
solution much more difficult.

\begin{figure}
\includegraphics[width=14cm,height=12cm]{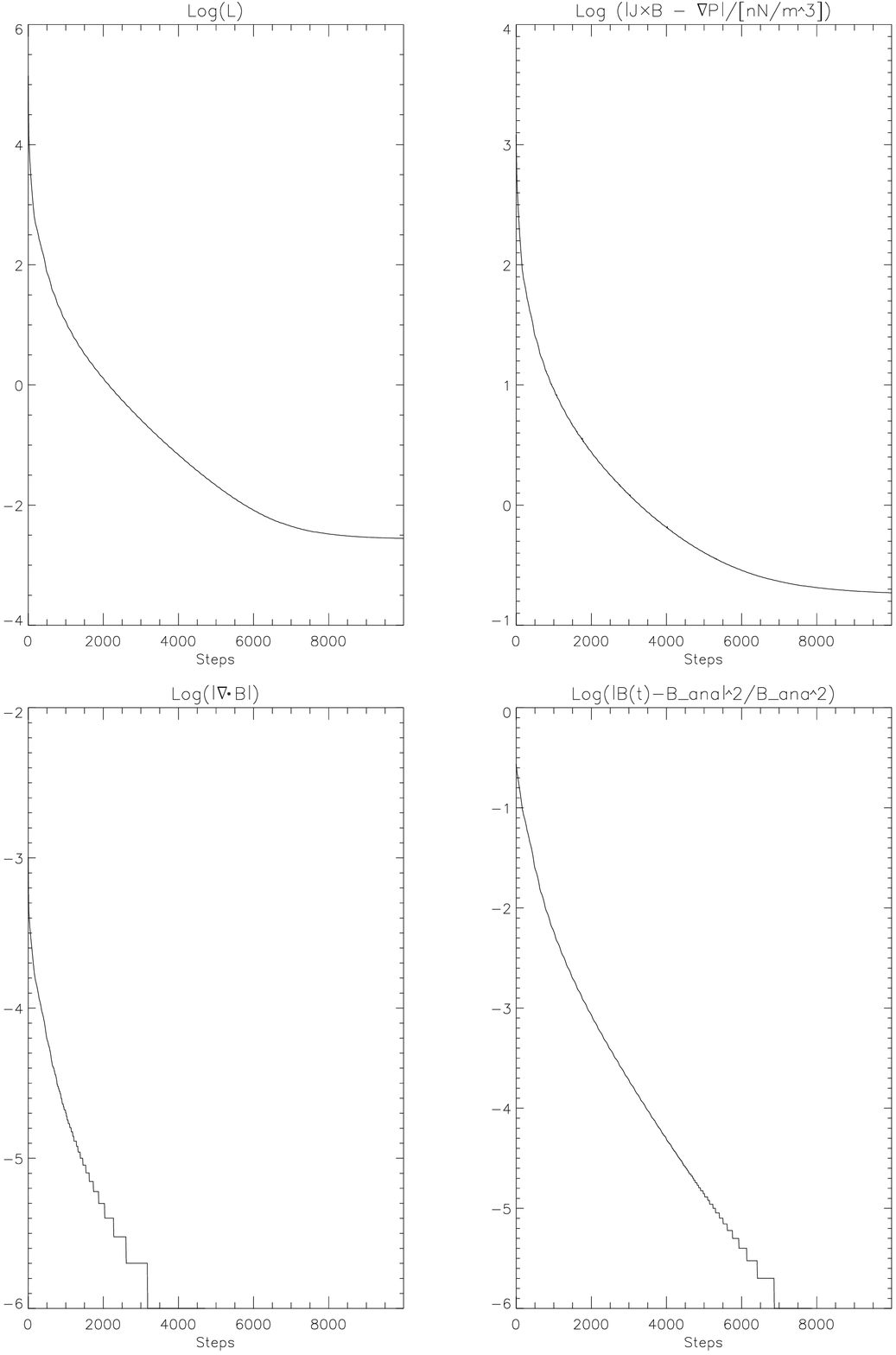}
\caption{MHS-1: Evolution of $L$, force balance
$|{\bf j} \times {\bf B} -\nabla P|$, $|\nabla \cdot {\bf B}|$
and the difference between the numerical magnetic field and the
known analytical solution
$|{\bf B}(t)- {\bf B_{\rm ana}}|^2 / |{\bf B_{\rm ana}}|^2 $.
All quantities are averaged over the numerical grid.}
\label{fig2}
\end{figure}

In figure \ref{fig1} we show three-dimensional plots of selected field lines
for this MHS-equilibrium. The colour coding of the bottom boundary indicates
the distribution of $B_n$ on that boundary. To test our code we try to
reconstruct that equilibrium in the following way:
\begin{itemize}
\item Inside the computational box we choose a potential magnetic field as
start equilibrium.
\item We prescribe the plasma pressure in the box with the analytic solution.
\item We prescribe the vector magnetic field on the boundaries of the
computational box.
\item We iterate for the magnetic field inside the computational box with
(\ref{iterateB}).
\end{itemize}
During the computations we calculate
the quantities $L$, the absolute
value of the force balance $|{\bf J} \times {\bf B}-\nabla P |$
(averaged over the numerical grid), the value
of $|\nabla \cdot{\bf B}|$ (averaged over the numerical grid),
and the difference
between the numerical magnetic field and the
known analytical solution
$|{\bf B}(t)- {\bf B}_{\rm ana}|^2 / |{\bf B}_{\rm ana}|^2  $
(averaged over the numerical grid) at each time step.
In figure \ref{fig2} we show the development of these quantities during
the iteration process with logarithmic scaling. All quantities decrease over
several orders of magnitude during the optimization process and reach the
level of the discretisation error. The discretisation error corresponds to
the value of $L\; $,  $|{\bf J} \times {\bf B}- \nabla P -\rho \; \nabla \Psi|$ and
$|\nabla \cdot{\bf B}|$ for the analytic solution computed on a numerical
grid. In the upper half of table \ref{resultstab} we summarize the main result
of the iteration process.
 The first row corresponds to the
analytic solution computed on a grid and defines the discretisation error.
The second row contains the values of $L$, the force balance and the relative
error for the start configuration, where the interior points have been
replaced by a potential field. The relatively large values of the three
 quantities are due to the deviation from the equilibrium.
 The next rows show how the three diagnostic quantities evolve during
 the iteration.
After 10000 iteration-steps the discretisation error is reached for all
values and the original equilibrium MHS-1 has been reconstructed.

We use a Landweber iteration (see e.g. \cite{louis}) with some
automatic control of the stepsize. The continuous
form of equation (\ref{iterateB}) ensures a monotonously decreasing $L$.
A monotonously decreasing $L$ in the discretized form is ensured if
the iteration step $dt$ is sufficiently small.
The code checks if
$L(t+dt) < L(t)$ after each time step and if the condition is not fulfilled,
the iteration step is refused and $dt$ is reduced by a factor of 2.
After each successful iteration step we increase dt slowly by a factor
of $1.01$ to allow the time step to become as large as possible with respect
to the stability condition. In principle there are more sophisticated methods
available to calculate an effective $dt$ for each iteration step
(see e.g. \cite{geiger}) but these methods have a huge numerical overhead
and further numerical experiments will have to show which of these are
favourable for our problem.

\begin{table}
\caption{Details of runs to reconstruct MHS equilibria.}
\begin{tabular}{|r|r|r|r|r|r|}
\hline
$n_x \times n_y \times n_z$ & Step & $\frac{\DS L}{\DS [{\rm T}^2 m]}$ &
$\frac{\DS \rm Force-balance}{\DS [{\rm nN \, m}^{-3}]}$& Relative Error \\
\hline
$40 \times 40 \times 20 $ &{\bf MHS-1}&$ 0.0028$&$ 0.19 $ & Reference \\
Start & 0&$140613 $&$1205$ &$0.27$ \\
$ $ & 500 &$74 $&$25.5$ &$0.025$ \\
$ $ & 5000 &$0.021 $&$0.4$ &$1.4 \; 10^{-5}$ \\
$ $ & 10000 &$0.0028 $&$0.19$ &$ <10^{-6}$ \\
\hline
%
 $40 \times 40 \times 200 $ &{\bf MHS-2}&$ 4.3$&$ 10.7 $ & Reference \\
Start & 0&$ 3.7\; 10^7$&$15290 $ & $ 0.14$ \\
$ $ & 500&$ 101423$&$2399 $ & $0.017 $ \\
$ $ & 1000&$3393$&$430 $ & $9 \; 10^{-4} $ \\
$ $ & 5000&$4.3$&$10.8 $ & $<10^{-6} $ \\
$ $ & &$ $&$ $ & $ $ \\
$ $ & &$ $&$ $ & $ $ \\
 \hline
\end{tabular}
\label{resultstab}
\end{table}

\subsection{Equilibrium MHS-2, Helmet Streamer}
\begin{figure}
\includegraphics[width=10cm,height=15cm]{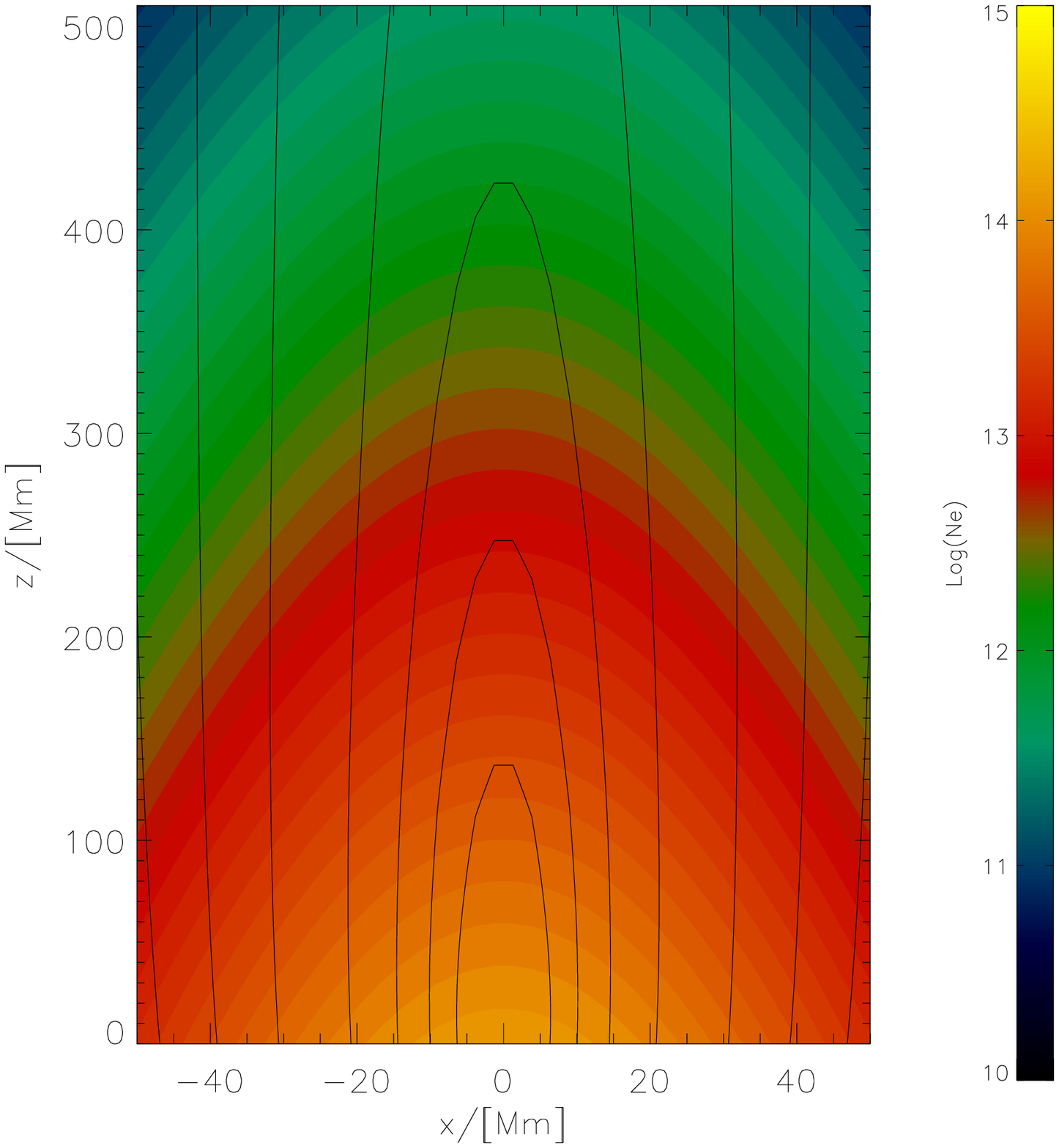}
\caption{MHS 2: A projection of some field lines for MHS-2. The back ground
colours correspond to the logarithm of the electron number density
$\frac{N}{[m^{3}]}$. Please note the different scale in x and z.}
\label{fig3}
\end{figure}
As a second example we consider an equilibrium with gravity which has been used to model
coronal helmet streamers \cite{twtnks1}. The method is based on the asymptotic expansion
method \cite{ks72} and corresponds to a nonlinear Grad-Shafranov equation.
Here we choose for the terms in (\ref{gse})
\begin{eqnarray}
P(A,\Psi)&=&a_0 \; \exp\left(-\frac{\Psi}{R T}\right) \; \exp(c \; A) \label{p2}, \\
B_y(A)&=&2 \; \sqrt{1-a_0} \;\exp\left(\frac{c}{2} \; A \right) \label{by2}.
\end{eqnarray}

For simplicity we use a constant gravity $\Psi=g \; z ,\;\;g=270 \frac{\mbox{m}}{\mbox{s}^2}$ and
a constant coronal temperature $T=3 \; 10^6 \,
\mbox{K}$ . Consequently we get the mass
density from the plasma pressure as $\rho=P/{RT}$. The parameters correspond
to a coronal pressure scale height
\begin{equation}
  H_0=\frac{k_B \; T}{m|\nabla \Psi|}=\frac{RT}{g}
\label{scaleheight}\end{equation}
of 93 Mm $\approx$ 0.13 solar radii. With help of the
method of asymptotic expansion we find an analytic solution of (\ref{gse}):
\begin{eqnarray}
A(x,z) &=&-\frac{2}{c} \; \log \left(\cosh \left(c \; \sqrt{\frac{p_0(z)}{2}}\; x \right)\right)
+\frac{1}{c} \; \log\left(\frac{p_0(z)}{k_0(z)}\right) \label{A2}, \\
p_0(z) &=& \frac{1}{1+\lambda \; z}, \\
k_0(z) &=& \exp\left(-\frac{\Psi}{RT} \right)= \exp\left(-\frac{z}{H_0}
\right).
\end{eqnarray}
We choose $c$ = 0.05 Mm$^{-1}$ and $\lambda$ = 0.001 Mm$^{-1}$ for the free
parameters and compute the solution on a grid of $nx$ = $ny$ = 40,
$nz$ = 200.
With help of the flux function (\ref{A2}), the equations for the pressure
(\ref{p2}), shear field (\ref{by2}) and the magnetic field definition (\ref{defB}) we get
the magnetic field {\bf B} and the plasma pressure $P$. For an isothermal
plasma we derive the number density as $N=\frac{P}{k_B \; T}$.
Let us remark that the quantity $N$ is
what we will get for real data from coronal measurements after the
tomographic reconstruction. In principle a non-constant temperature
$T$ can also be used, e.g. from a standard atmosphere model.
Figure \ref{fig3} shows a projection of magnetic field lines and
the electron density structure as background for the helmet streamer
configuration MHS-2.

The solution is invariant in $y$ and we rotate the solution around the
z-axis with an angle $\phi=\pi/10$ which is useful for testing our
reconstruction code (All derivatives appear).
As start magnetic field we choose a Harris-sheet,
where the magnetic field has only one component
$B_z=-\sqrt{2} \; \tanh\left(\frac{c}{\sqrt{2}} \; x \right)$ which is invariant
in $z$ and $y$.
\begin{figure}
\includegraphics[width=14cm,height=12cm]{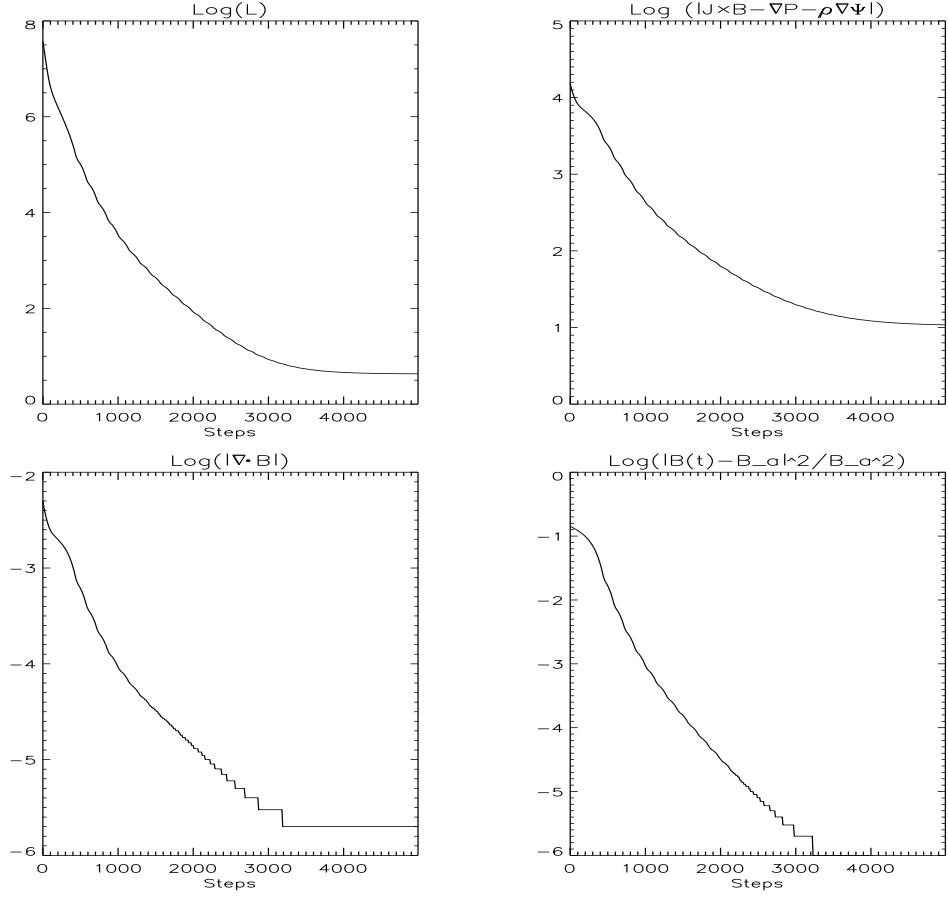}
\caption{MHS-2: Evolution of $L$, force balance
$|{\bf j} \times {\bf B} -\nabla P- \rho \nabla \Psi|$, $|\nabla \cdot {\bf B}|$
and the difference between the numerical magnetic field and the
known analytical solution
$|{\bf B}(t)- {\bf B_{\rm ana}}|^2 / |{\bf B_{\rm ana}}|^2 $.
All quantities are averaged over the numerical grid.}
\label{fig4}
\end{figure}
We apply our code for the reconstruction of this helmet streamer equilibrium
MHS-2 in the following way:
\begin{itemize}
\item Inside the computational box we choose a Harris-sheet magnetic field as
start equilibrium.
\item We prescribe the electron number density $N$ in the box with the analytic solution.
\item Under the assumption of a constant coronal temperature and a constant
gravity we calculate the plasma pressure, density and compute
${\bf u} =-\mu_0 (\nabla P +\rho \nabla \Psi)$ on
the grid.
\item We prescribe the vector magnetic field  on the boundaries of the
computational box. Similar as in the previous example we use the analytic
solution to fix the magnetic field on all boundaries for simplicity.
\item We iterate for the magnetic field inside the computational box with
(\ref{iterateB}).
\end{itemize}
Similar as for MHS-1 we diagnose
the quantities $L$, the absolute
value of the force balance $|{\bf J} \times {\bf B}-\nabla P -\rho \; \nabla \Psi|$
(averaged over the numerical grid), the value
of $|\nabla \cdot{\bf B}|$ (averaged over the numerical grid),
and the difference
between the numerical magnetic field and the
known analytical solution
$|{\bf B}(t)- {\bf B}_{\rm ana}|^2 / |{\bf B}_{\rm ana}|^2 $
(averaged over the numerical grid) at each time step.
In figure \ref{fig4} we show the development of these quantities during
the iteration process with logarithmic scaling. All quantities decrease over
several orders of magnitude during the optimization process and reach the
level of the discretisation error. The discretisation error corresponds to
the value of $L\; $,  $|{\bf J} \times {\bf B}- \nabla P -\rho \; \nabla \Psi|$ and
$|\nabla \cdot{\bf B}|$ for the analytic solution computed on a numerical
grid. Please note that the discretisation error for MHS-2 is significantly
larger as for MHS-1 due to the nature of nonlinear analytic solution.
In the lower half of table \ref{resultstab} we summarize the main result
of the iteration process.

The first row corresponds to the
analytic solution computed on a grid and defines the discretisation error.
The second row contains the values of $L$, the force balance and the relative
error for the start configuration, where the interior points have been
replaced by a Harris-sheet magnetic field. The large values of the three
 quantities are due to the deviation from the equilibrium.
 The next rows show how the three diagnostic quantities evolve during
 the iteration.
After 5000 iteration-steps the discretisation error is reached for all
values and the original helmet streamer-configuration MHS-2 has been reconstructed.
\subsection{force-free and MHS reconstruction for different $\beta$}
\label{plasmabeta}
It is generally assumed that the magnetic pressure in the lower corona
is much larger than the plasma pressure leading to $\beta \ll 1$.
For longer structures like helmet streamers the plasma $\beta$
increases significantly. It is generally assumed that the effects of plasma
pressure and gravity can be neglected for low $\beta$ plasmas leading to a
nearly force-free state. Let us remark that one can construct high $\beta$
force-free equilibria by adding a homogeneous plasma pressure
$\nabla P=0$ (or a barometric density distribution
$\nabla P=-\rho \; \nabla \Psi $ for configurations with gravity)
to any exact force-free configuration. Here we do not study
such singular cases, but more realistic configurations where
$\nabla P \propto \frac{P}{l}$, where $l$ is a typical length scale of the
problem. The equilibrium MHS-1 allows us to compute configurations with
different plasma $\beta$ by prescribing the parameter $a_0$, where $a_0=0$
corresponds to an exact force-free state. We use our code to
investigate how well a magnetic field configuration can be reconstructed by
the force-free approach. We start all
reconstruction runs with a potential field solution $c=0.0$ and
the magnetic field boundary conditions extracted from the exact solution
(artificial vector magnetograms). The configurations in the left hand side
in table \ref{beta}
have been reconstructed by the assumption of a force-free configuration and
the configurations in the right hand side in table \ref{beta} are
reconstructed as MHS-equilibria. The force-free reconstruction needs the
boundary magnetic field data as input and the MHS-reconstruction additional
requires the plasma-density structure, which we extract here
from the analytical solution.
This corresponds to artificial tomographic
information.

We diagnose the quantities $L$
(${\rm L_{FF}}$ for ForceFree and ${\rm L_{MHS}}$ for MagnetoHydroStatic)
and the deviation from the the analytic solution
(${\rm Error_{FF}}$ for ForceFree and ${\rm Error_{MHS}}$ for MagnetoHydroStatic)
similarly as described in the previous sections.
 For MHS-reconstruction our code finds the magnetic field
structure for all configurations with an error corresponding to the discretisation
error. If we restrict our code to an exact force-free reconstruction we still
get a considerable good agreement with the exact solution for a plasma
$\beta$ of less than $10^{-3}$. For higher values of $\beta$ both the
value of $L$ (where $L=0$ corresponds to an exact force-free state) and the
error in the magnetic field increases. Consequently our code finds the
expected result that the effect of plasma pressure is neglible  for low $\beta$
configurations. The result also shows that a direct consideration of
tomographic information regarding the electron density is only useful for
finite $\beta$ plasmas. For low $\beta$ plasmas the magnetic field structure
is practically not influenced by the plasma density distribution.
Let us remark that it is still possible to consider
some indirect information provided by the plasma density for low $\beta$
plasmas, e.g. the fact that the density gradient parallel to the magnetic
field is much lower than the density gradient perpendicular to the magnetic
field. Consequently the magnetic field lines are outlined by the emitting
plasma. This allows to consider stereoscopic information for the reconstruction
of low $\beta$ plasmas \cite{twtn02}.

\subsection{Speed of the method}
The speed of our code is approximately proportional to $N^5$ (N is the
number of points for one side of the computational box) similar as found by
\cite{wheatland00} for the force-free case. This $N^5$ dependence looks
discouraging for the reconstruction of large boxes. We undertake
some rough estimations if the method is practical for modern vector
magnetographs. For a grid of N=40 a reconstruction takes about 5 min on a 4
processor computer. The IVM vector magnetograph in Hawaii has a resolution of
N=256 pixel. Consequently a reconstruction with full IVM-resolution would
take approximately $5 \mbox{min} \cdot (256/40)^5 \approx 35 \mbox{days}$. If only the half IVM-resolution
is used the reconstruction time would be $5 \mbox{min} \cdot (128/40)^5 \approx 28 \mbox{hours}$, which
seems to be acceptable. We are optimistic that an improved numerical scheme
(e.g. using conjugated gradients or multi-grid methods), a massive
parallelization (Using 16-32 processors instead of 4.\footnote{The method seems
to parallelize quite well. On 4 processors the reconstruction is about 3 times faster
than on one processor.}) and increasing computer
speed will speed up the reconstruction time significantly. The computing time
for the optimization approach seems to be high, but comparisons of the
optimization method with classical MHD relaxation (for the force-free case)
have shown that the
optimization method is more effective \cite{twtn03}. Direct
extrapolation methods (e.g. \cite{wu85}) are much faster than iterative
methods but known to become unstable with increasing coronal height.
\begin{table}
\caption{force-free and MHS-reconstruction. 
The first column contains the plasma $\beta$, the second column $a_0$, the third
column the final value of $L$ for a force-free reconstruction, the fourth column the
error in the magnetic field structure compared to the analytic solution, the fifth column
the final $L$-value for a MHS-reconstruction and finally the sixth column the
error in the magnetic field structure compared to the analytic solution for the
MHS-reconstruction. All runs have been computed with MHS-1 on a grid $n_x=n_y=40, \; n_z=20$
for $5000$ iteration steps.}
\begin{tabular}{|r|r|r|r|r|r|}
\hline
Plasma $\beta$ & $a_0$ & ${\rm L_{FF}}$ & ${\rm Error_{FF}}$ & ${\rm L_{MHS}}$ & ${\rm Error_{MHS}}$  \\
\hline
$ 0$      & $ 0 $        & $0.027$ & $1.5 \;10^{-5} $ && \\
$10^{-4}$ & $ 0.00045  $ & $0.028$ & $1.5 \;10^{-5} $ & $0.028$ & $1.5 \;10^{-5} $\\
$10^{-3}$ & $ 0.0045  $ & $0.04$ & $1.5 \;10^{-4} $ & $0.028$ & $1.5 \;10^{-5} $\\
$10^{-2}$ & $ 0.045  $ & $1.26$ & $4.9 \;10^{-4} $ & $0.028$ & $1.5 \;10^{-5} $\\
$10^{-1}$ & $ 0.41  $ & $112$ & $6.0 \;10^{-3} $ & $0.027$ & $1.7 \;10^{-5} $\\
$0.2$ & $ 0.74  $ & $807$ & $ 0.05 $ & $0.015$ & $1.0 \;10^{-5} $\\
$0.3$ & $ 1.0  $ & $1443$ & $ 0.2 $ & $0.007$ & $2.0 \;10^{-6} $\\
\hline
\end{tabular}
\label{beta}
\end{table}
%
\section{Using coronal magnetic field information as a constraint for tomography}
\label{tomoopt}
Until now, we used information regarding the coronal density and pressure
structure as given.  In principal we could consider density and pressure in
functional (\ref{defL}) as additional variables to be optimized just as $\mathbf
{B}$. In this case, however, the problem of minimizing $L$ would be hopelessly
underdetermined even if the boundary values of $\mathbf {B}$, $\rho$ and $P$
were given at the Sun's surface.

If the magnetic field was known on the other hand, we could use (\ref{defL}) and
assume a temperature variation to determine a consistent density and pressure.
Even though we will find immediately that this approach is doomed to fail
we here mention for completeness the optimization equations which can be derived
from (\ref{defL})
\begin{eqnarray*}
 \rho = m N,  \quad P = k_B T N
\end{eqnarray*}
and we get (see the appendix  \ref{appendixB})
\begin{eqnarray}
\frac{1}{2} \; \frac{d L}{d t} &=&
\int_{V} H \;  \frac{\partial N}{\partial t}   \; d^3x
+\int_{S} I \; \frac{\partial N}{\partial t} \; d^2x
\label{minimize2}, \\
H &=& \mu_0 m \; {\bf \Omega_a} \cdot \nabla \Psi
-\mu_0 \; k_B T \; \nabla \cdot {\bf \Omega_a}, \\
I &=&  \mu_0 \; k_B T \; {\bf \Omega_a} \cdot {\bf \hat n}.
\end{eqnarray}
(See (\ref{defomega}) for the definition of ${\bf \Omega_a}$ and ${\bf \Omega_b}$.)
\footnote{
Let us remark that the general form of $\frac{d L}{d t}$ will vary both
the density distribution and the magnetic field
$$
\frac{1}{2} \; \frac{d L}{d t} =
-\int_{V} \frac{\partial {\bf B}}{\partial t} \cdot {\bf F}
+ H \;  \frac{\partial N}{\partial t} \; d^3x
-\int_{S} \frac{\partial {\bf B}}{\partial t} \cdot {\bf G}
+I \; \frac{\partial N}{\partial t} \; d^2x.
$$
}
Physically speaking, the solution to these equations yield a
pressure which exactly balances excessive magnetic field forces onto the
plasma. However, due to the small value of $\beta$ in the corona,
already a small relative error in the assumed magnetic field will
result in residual forces which need a huge plasma pressure to be balanced so that
a small relative error in $\mathbf{B}$ will produce a large relative
error in $N$. We obviously need another approach to get hold of a decent estimate of
the coronal density distribution and this must be is based on additional observations.

The coronal electron density can be observed by coronagraphs. Unfortunately,
coronagraphs yield only integrated column densities along the line-of-sight
and the 3D density distribution itself has to be reconstructed from these
measurements by means of a tomographic inversion
\cite{zidowitz99,frazin00,frazin02}.
This inversion process has besides its intrinsic ill-posednes  additional
problems to cope with if applied to coronagraph data:\\
$\bullet$ non-stationarity of the coronal density structures,\\
$\bullet$ incomplete data due to the occultation of the image centers
(exterior tomography problem),\\
$\bullet$ non-ideal viewing geometry due to the tilt of the Sun's axis
with respect to the ecliptic.\\
As a result of all these problems, the spatial resolution which ultimately
can be achieved with the reconstruction is limited.
The conventional procedure to obtain a reliable solution is to minimize
an expression of the following form:
\begin{equation}
G(N)=\sum\limits_{p,i}
  |I_{p,i}^{\rm obs}-{\cal I}_{p,i}(N)|^2
  +\mu \int\limits_V |R(N)|^2\, d^3x
\label{tomoeq},
\end{equation}
where $I^{\rm obs}_{p,i}$ is the observed intensity in pixel $p$ of image $i$
and ${\cal I}_{p,i}$ is the respective simulated intensity which is
calculated from a density $N$ as a line-of-sight integral
\begin{equation}
 {\cal I}_{p,i}=\int\limits_{{\cal C}_{p,i}} N d\ell
\label{losinteg}.
\end{equation}
Here, ${\cal C}_{p,i}$ is the beam from pixel $p$ of image $i$, i.e. the location
of points $\in$ V which project onto the respective pixel.
We omit here modifications of (\ref{losinteg}) due to the scattering geometry
and the scattering crossection of the electrons which lead to slight
variations of the integrand in (\ref{losinteg}).
In (\ref{tomoeq}), $R$ is a regularization function to be specified below.
The primary aim is to
find a density model $N$ which makes the first term vanish. In this case $N$
is compatible with the observations.  However, due to measurements errors,
inconsistencies of the observations mentioned above and its possible
insensitivity to certain density structures, it does not make sense to minimize
the first term alone below a level approximately given by the measurement error
variance.  To stabilize the model reconstruction on density structures to which
the observations are insensitive, the regularization term is added with a
regularization parameter $\mu$ tuned so that the first term is approximately
brought down to the observation error variance when the complete expression $G$
is minimal.

Since ${\cal I}(N)$ is basically an integration operator,
$R(N)$ very often is chosen as a differential operator, e.g.,
\begin{equation}
R(N) = \nabla^2 N
\label{diffreg}
\end{equation}
As a result, while the integration makes ${\cal I}(N)$
insensitive to small scale structures, $R(N)$ responds to
these but has little effect on the large scales. Therefore, when minimizing
$G$, the large scales are shaped by the first term to comply with the
observations, while the small scales are kept smooth by the regularization
term in $G$. The transition between large and small scales is mainly
determined by the weight $\mu$ of the regularization in $G$
which must be enhanced the worse the quality of the observations.
In this sense, the spatial resolution of the model density $N$
which can be achieved depends largely on errors, inconsistencies and gaps
of the data.

But the power of the regularization term goes beyond suppressing unwanted small
scale noise in the reconstructed model $N$. Any additional physical constraint
for $N$ can be included here just like the
$\nabla\cdot\mathbf{B}=0$ constraint was added to the force balance condition in
(\ref{defL}).
One obvious constraint for the density is its positivity, which can be taken
account of in $R$ by so-called barrier functions (see \cite{frazin00}).
For a more stabilizing constraint we may return to (\ref{minimize2}).
It was derived so as to minimize the square of $\mathbf{\Omega_a}$ which
is proportional to the local force balance. Note that $\mathbf{\Omega_b}$
has no dependence on $N$ and therefore is not varied here. The argument
which led us to discard (\ref{minimize2}) as a starting point for an iteration
for the density mainly applies to the force components
$\mathbf{\Omega_a}_{\perp}$ $\propto$
$(\nabla\times\mathbf{B})\times\mathbf{B}$
$-$ $\mu_0 (\nabla_{\perp}P +\rho\nabla_{\perp}\Psi)$
perpendicular the magnetic field which are
dominated by the magnetic term to lowest order.
We therefore choose a regularization term which
at least takes care of the field-aligned force balance in
$ \Omega_{a,\parallel}$ $\propto$
$-$ $\mu_0 (\nabla_{\parallel}P +\rho\nabla_{\parallel}\Psi)$.
This is achieved by
\begin{equation}
 R(N) = \frac{1}{k_B T} (\hat{\mathbf{B}}\cdot\nabla) P
     + \rho(\hat{\mathbf{B}}\cdot\nabla) \Psi
      = (\hat{\mathbf{B}}\cdot\nabla) N
      - \frac{\hat{\mathbf{B}}\cdot\hat{\mathbf{r}}}{H_0} N
\label{bfldreg},
\end{equation}
where again an isothermal plasma is assumed and
$H_0$ is the pressure scale height of the solar corona as in
(\ref{scaleheight}).
An extension to a varying temperature is straight forward if it is given,
however we cannot solve for an unknown $T$ unless we make use of additional
observations.

The regularization term (\ref{bfldreg}) effectively smoothes the density
out along the magnetic field lines and thereby takes account of the fact
that the transport coefficients in a magnetized plasma are much larger along
the magnetic field than in perpendicular direction.
In perpendicular direction to $\hat{\mathbf{B}}$ the density may have
gradients as sharp as our model resolution allows without
changing the value of $G$.

\begin{figure}
\hspace*{\fill}
\includegraphics[bb=67 176 556 671,clip,height=6.8cm,angle=-90]
{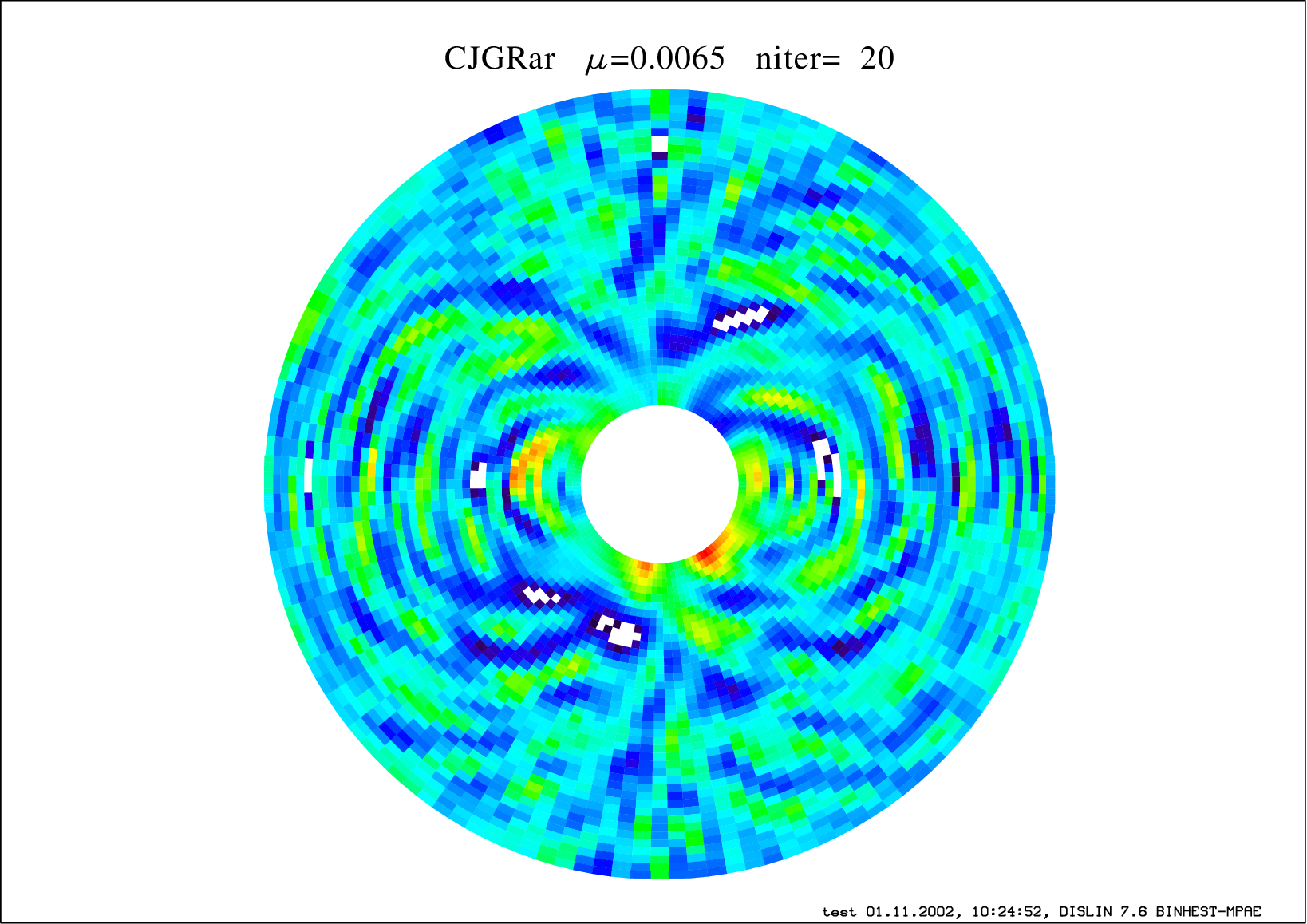}
\hspace*{\fill}
\caption{A reconstruction of completely noisy images
with the regularization term (\ref{bfldreg}). The visible structures are
imprints of the dipole magnetic field assumed in the regularization term.}
\label{Fig:noisemod}
\end{figure}

The magnetic field assumed in the test calculations below was a simple
dipole field. The effect of the above regularization operator becomes
particularly visible if instead of meaningful data, we assume that
$I^{\rm obs}$ is pure noise. The magnetic field then is the only real
information in the inversion process and it becomes directly visible in the
reconstruction results (see Fig~\ref{Fig:noisemod}).

In appendix \ref{appendixC} we derive an expression for an iterative descent
step analogous to (\ref{minimize2}) but which is preferable to (\ref{minimize2})
because it includes the additional observations to stabilize the reconstruction
\begin{equation}
\Derv{G}{t}=\int\limits_V H \Dpar{N}{t} d^3x
           +\int\limits_S I \Dpar{N}{t} d^2x
\label{minimize3},
\end{equation}
where
\begin{eqnarray}
H(\mathbf{x}) &=& 2\sum\limits_{p,i} \delta_{{\cal C}_{p,i}}(\mathbf{x})
                ({\cal I}_{p,i}(N)-I_{p,i}^{\rm obs})
\nonumber \\
 &-&2\mu \big[(\hat{\mathbf{B}}\cdot\nabla)R(N)
 + \frac{\hat{\mathbf{B}}\cdot\hat{\mathbf{r}}}{H_0} R(N)\big]
\label{minimize3H}, \\
I(\mathbf{x}) &=&
 2\mu (\hat{\mathbf{B}}\cdot\hat{\mathbf{n}}) R(N)
\label{minimize3I}
\end{eqnarray}
and $\delta_{{\cal C}_{p,i}}$ is 1 inside the beam emanating from pixel $p$ of
image $i$ and 0 otherwise (see appendix \ref{appendixC}).

In order to test the feasibility of our scheme, we minimize (\ref{minimize3})
for a 2D test model by means of a conjugate gradient iteration.
In table \ref{Tab:regtest} we compare the action of this operator with more
conventional means of regularization for a two-dimensional reconstruction.
Here, the model and data errors of a reconstructed model density $N$ are
defined as
\begin{eqnarray}
  \mathtext{data error} &=& \frac{1}{2}
  \sum\limits_{p,i}\abs{I_{p,i}^{\rm obs}-{\cal I}_{p,i}(N)}^2,
\nonumber \\
  \mathtext{model error} &=& \frac{1}{2} \int\limits_V
                \abs{N_{\rm ana}-N}^2\, d^3x.
\nonumber
\end{eqnarray}
Here, $N_{\rm ana}$ is the analytic density model used to obtain the data
$I^{\rm obs}$.
The reconstruction a) was obtained without explicit regularization (i.e. with $\mu$ = 0).
Instead, the iteration was stopped after 13 iterations when the model error
was minimal. Subsequent iterations further decrease the data error but
enhance the model error (which for real data is not known), a phenomenon
which is known as semiconvergence.
For reconstructions b) and d) the conventional regularization operator
(\ref{diffreg}), for c) and e) the operator (\ref{bfldreg}) was used,
however with the scale height $H_0$ set to $\infty$.

In cases d) and e), the regularization parameter $\mu$ and the number of
iterations where optimized to achieve the best agreement with the original
model. We show these results only to demonstrate how close a reconstruction
can come to the true solution in principle.
In practical cases, however, the true density distribution is not known
and the optimum solution has to be sought based only on the values for the data
error and the regularization term. In cases b) and c) these values have been
used in an L-curve search for the optimum solution (Hansen and O'Leary, 1993).

\begin{figure}
\hspace*{\fill}
\begin{picture}(12.0,12.0)
\put( 0.0,12.0){\includegraphics[bb=67 176 556 671,clip,height=6cm,angle=-90]
{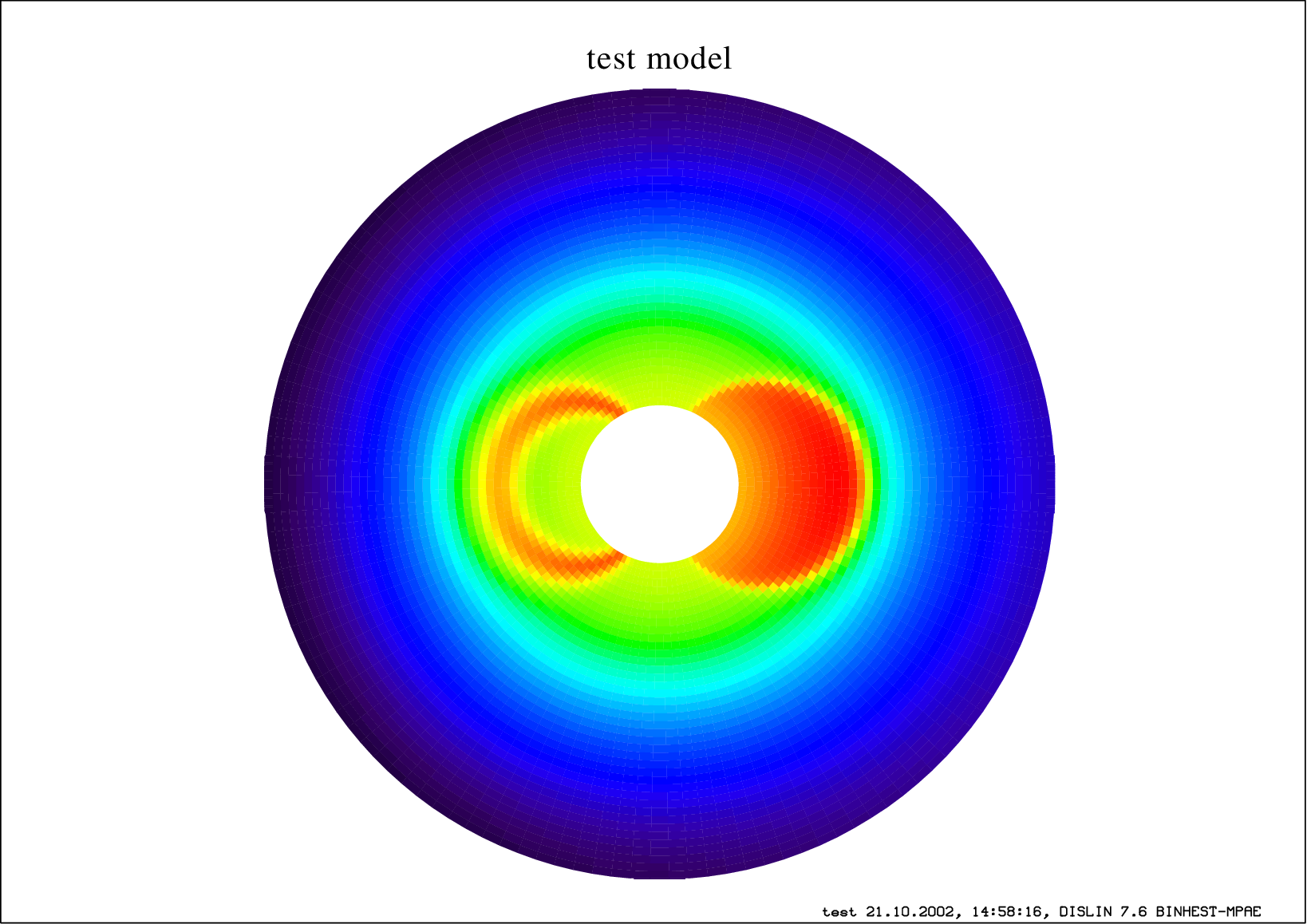}}
\put( 3.0, 9.0){\makebox(0,0){original}}
\put( 6.0,12.0){\includegraphics[bb=67 176 556 671,clip,height=6cm,angle=-90]
{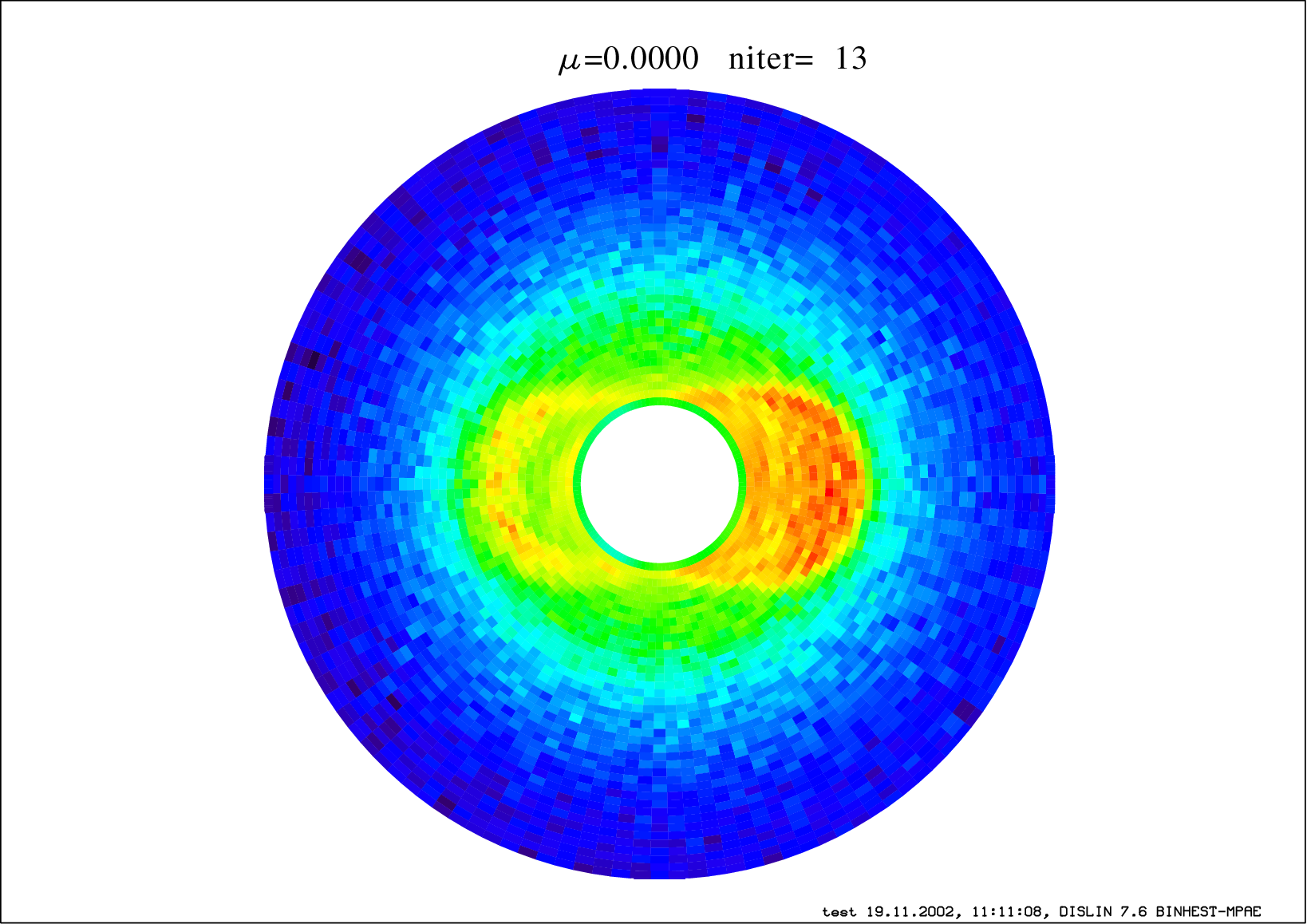}}
\put( 9.0, 9.0){\makebox(0,0){a)}}
\put( 0.0, 6.0){\includegraphics[bb=67 176 556 671,clip,height=6cm,angle=-90]
{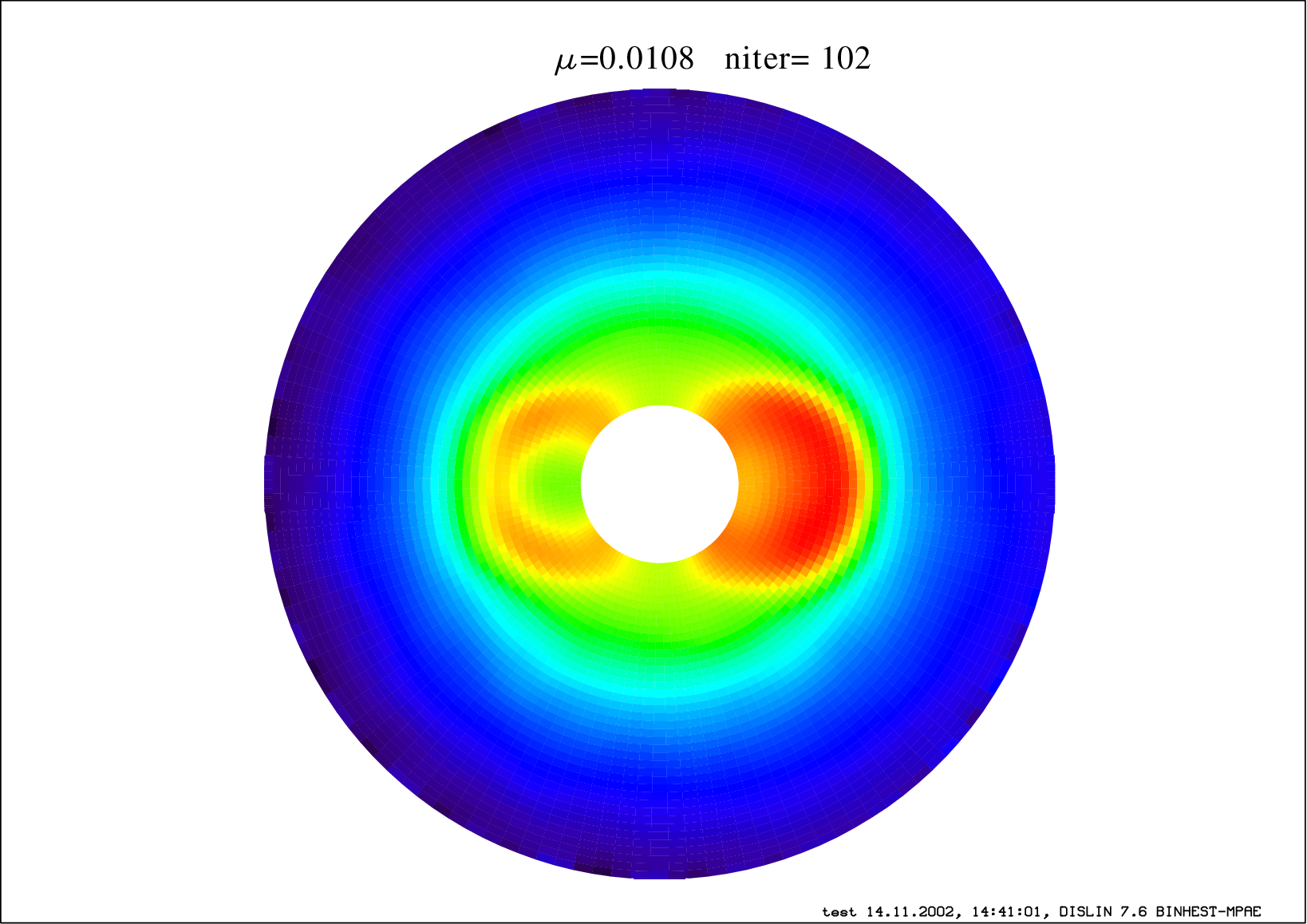}}
\put( 3.0, 3.0){\makebox(0,0){b)}}
\put( 6.0, 6.0){\includegraphics[bb=67 176 556 671,clip,height=6cm,angle=-90]
{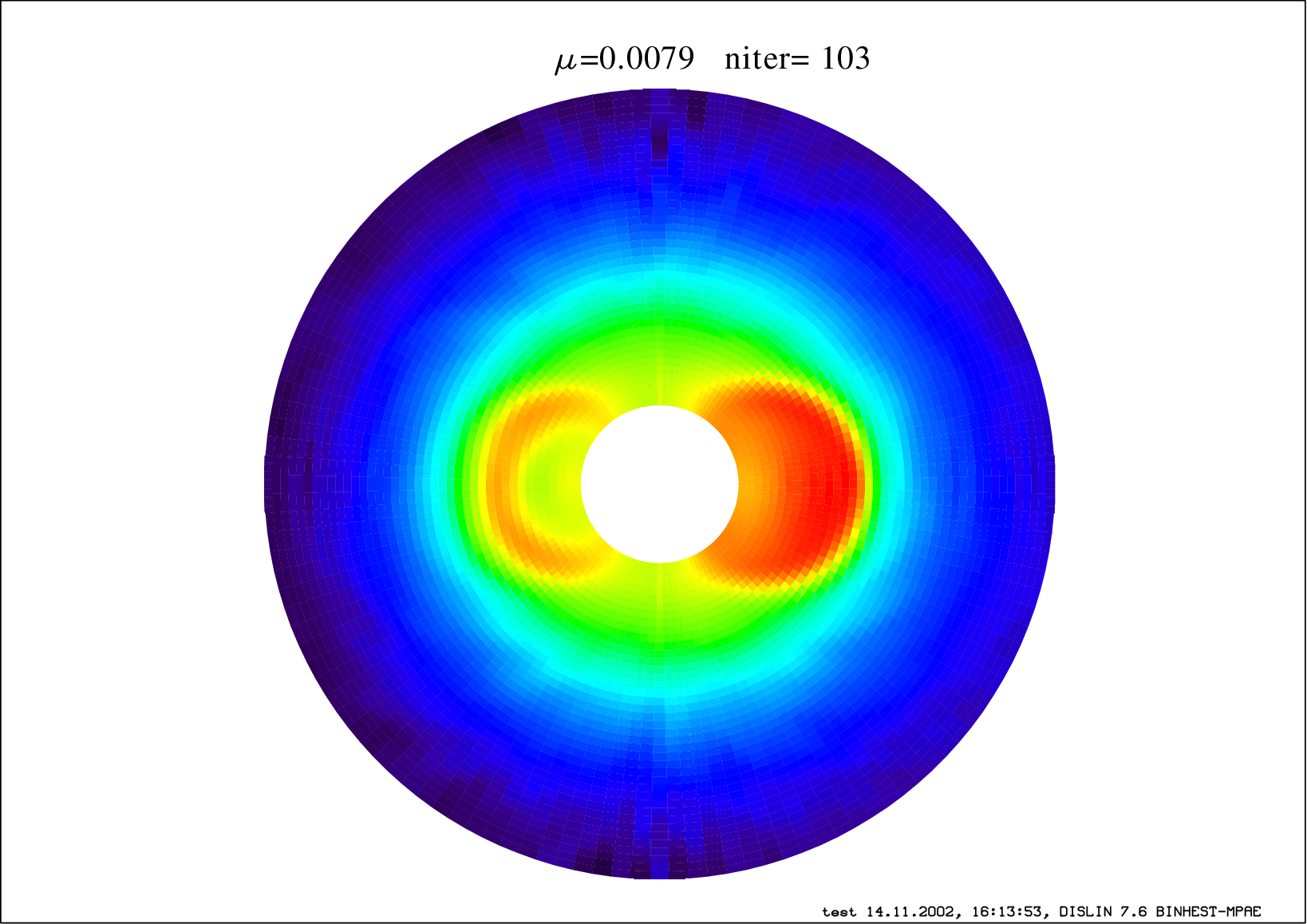}}
\put( 9.0, 3.0){\makebox(0,0){c)}}
\end{picture}
\hspace*{\fill}
\caption{Comparison of different regularization
methods for the reconstruction of the original model $N_{\rm ana}$
in the upper left. The results a) to c) correspond to the cases in
table \ref{Tab:regtest}.}
\label{Fig:regtest}
\end{figure}

In Fig.~\ref{Fig:regtest} we show the models associated with the test
inversions in table \ref{Tab:regtest}.
The upper left shows the original model simulating a coronal loop on the
western limb and a streamer on the eastern limb.
This model was used to calculate the artificial data used as input for the
reconstruction after some noise was
added. The standard deviation of the noise was 3$\cdot 10^{-2}$ times the
square root of the local data intensity when the maximum data intensity
is normalized to unity.

\begin{table}
\begin{tabular}{|l|c|c|c|} \hline
  Regularization method & optimal iteration & model error & data error \\[-2mm]
                        & steps &      &  \\ \hline
  a) stopping rule, $R$=0                & 14    & 12.4 & 6.46 10$^{-5}$\\ \hline
  b) $R$=(\ref{diffreg}), $\mu$=0.01082  & 102   & 1.65 & 13.9 10$^{-5}$\\
  c) $R$=(\ref{bfldreg}), $\mu$=0.00787  & 103   & 1.21 & 12.5 10$^{-5}$\\ \hline
  d) $R$=(\ref{diffreg}), $\mu$=0.0050   & 53    & 1.21 & 12.2 10$^{-5}$\\
  e) $R$=(\ref{bfldreg}), $\mu$=0.0065   & 120   & 1.13 & 12.0 10$^{-5}$\\ \hline
\end{tabular}
\caption{Comparison of different regularization
methods. The model and data errors are defined in the text.
The results a) to c) correspond to the cases in Fig. \ref{Fig:regtest}.
Case a) uses no regularization at all and has a minimum model error
after 14 iteration steps.
Cases b) and d) use the isotropic regularzation as in (\ref{diffreg}),
cases c) and e) the isotropic regularzation as in (\ref{diffreg}).
Iteration steps and $\mu$ are optimized in b) and c) from an L-curve
search, in cases d) and e) for a minimum model error.)
}
\label{Tab:regtest}
\end{table}

In terms of the model error, (\ref{bfldreg}) yields slightly better results
than (\ref{diffreg}). The major improvement comes from the region close to the
occulter. In this region conventional tomography can only yield a limited
resolution because close to the occulter the structures are only seen in few
observations. On the other hand, this is the area where we observe the
strongest gradients in the density structures and where spatial resolution is
needed most. This fundamental lack of resolution can only be overcome if new
information (either observations or assumptions) is fed into the inversion
process. The new regularization operator (\ref{bfldreg}) contains this additional
information in form of the local magnetic field direction.

The price we have to pay is a more lengthy computation as the number of
iterations increases (see Table \ref{Tab:regtest}). Another problem which might
occur are spurious field-aligned density structures in the reconstruction
which add up to zero in the tomographic projections. Formally, (\ref{bfldreg})
has a nullspace which asymptotically comes close to the nullspace of the
${\cal I}$. In practical cases, however, the discretization error in the differentiation
is sufficient to give (\ref{bfldreg}) some isotropic component so that
even exactly field-aligned structures yield a small non-zero contribution
in a discretized operator (\ref{bfldreg}).

\section{Conclusions}
\label{conclusions}
In this paper we undertook a first step towards the inclusion of tomographic
information into the reconstruction of coronal magnetic fields and a first
step towards the inclusion of coronal magnetic field information into the
tomographic inversion procedure.
Until now we considered the reconstruction of the magnetic field from
its boundary values with an optimization code when the density structure is
given and the tomographic reconstruction of the coronal density distribution
from coronagraph data under the constraint of a given magnetic field.
As neither $N$ and nor $\mathbf B$ are known a priori in the solar corona,
we rather have to find a way to consistently reconstruct both quantities
from the observations simultaneously without the assumption that one of the
quantities is given.

In fact we observe that the two approaches discussed in the previous sections
are not only formally closely related but can be derived from a single
variational problem if the expression for $L$
is slightly modified and
the factor $B^{-2}$ in the integrand is omitted
\begin{eqnarray}
 L(\mathbf{B},N) &=&
        \int_V        \abs{(\nabla\times\mathbf{B})\times\mathbf{B}
                -\mu_0(\nabla P+\rho\nabla\Psi)}^2
                +B^2\abs{\nabla\cdot\mathbf{B}}^2
                        ]\;d^3x \nonumber \\
 &+& \frac{1}{\mu'}\sum\limits_{p,i}
      \abs{I_{p,i}^{\rm obs}-{\cal{I}}_{p,i}(N)}^2.
\label{BNvariation}
\end{eqnarray}
A variation with respect to $\mathbf{B}$ obviously leads to an iteration
scheme similar to (\ref{minimize1}), except for the effect of the $B^{-2}$
term in the integrand. We have tested the resulting scheme and found that it
also converges towards the analytic solution from which the boundary conditions
were taken, but the convergence speed was much slower than with
(\ref{minimize1}).
If $L$ is varied with respect to $N$ alone, we can ignore the terms which
depend only on $\mathbf{B}$ and we obtain (\ref{minimize3}) again if we discard the
perpendicular force balance in $\mathbf{\Omega_a}$.
This suggests that the individual reconstruction problems for $\mathbf{B}$
and $N$ are just two projections of a unique reconstruction problem.
In this case we could apply both algorithms simultaneously and
replace $N$ in the algorithm for $\mathbf{B}$ and $\mathbf{B}$ in the
algorithm for $N$ by the respective iterate and the problem as a whole should
converge as they do individually.
A test of this hypothesis will be attempted in the future.
The code is planed for use within the STEREO mission.
\acknowledgements
The authors thank Thomas Neukirch for useful discussions.
This work was supported by  DLR-grant 50 OC 0007.
We thank an unknown referee for useful comments.
\appendix
\renewcommand{\theequation}{\Alph{section}.\arabic{equation}}
\section{Derivation of {\bf F} and {\bf G} in (\ref{minimize1}).}
\label{appendixA}
\setcounter{equation}{0}
(\ref{defL}) can also be written as
\begin{equation}
L=\int_{V} B^2 \; (\Omega_a^2+\Omega_b^2) \; d^3x.
\end{equation}
\begin{eqnarray}
{\bf \Omega_a} &=& B^{-2} \;\left[(\nabla \times {\bf B}) \times {\bf B} + {\bf u} \right]
\nonumber\\
{\bf \Omega_b} &=& B^{-2} \;\left[(\nabla \cdot {\bf B}) \; {\bf B} \right]
\nonumber\\
{\bf u} &=&-\mu_0 (\nabla P +\rho \nabla \Psi)
\end{eqnarray}
We vary $L$ with respect to an iteration parameter $t$ and get
\begin{eqnarray}
\frac{1}{2} \; \frac{d L}{d t} &=&
\int_{V} {\bf \Omega_a} \cdot \frac{\partial}{\partial t}
[(\nabla \times {\bf B}) \times {\bf B} + {\bf u} ]  \; d^3x
\nonumber\\
&+&\int_{V} {\bf \Omega_b} \cdot  \frac{\partial}{\partial t}
[(\nabla \cdot {\bf B}) \; {\bf B}] \; d^3x
\nonumber\\
&-& \int_{V}  (\Omega_a^2 + \Omega_b^2) \;
{\bf B} \cdot \frac{\partial {\bf B}}{\partial t} \; d^3x.
\end{eqnarray}
Our aim is now to use vector identities and Gauss law in such way that all
terms contain a product with $\frac{\partial {\bf B}}{\partial t}$. This will
allow us to provide explicit evolution equations for ${\bf B}$ to minimize $L$.
The third term has the correct form already. We expand the first and second term
\begin{eqnarray}
\Rightarrow \frac{1}{2} \; \frac{d L}{d t} &=&
\int_{V} {\bf \Omega_a} \cdot \left[ \left(\nabla \times \frac{\partial {\bf B}}{\partial t}\right) \times {\bf B} \right]  \; d^3x
\nonumber\\
&+& \int_{V} {\bf \Omega_a} \cdot
\left[(\nabla \times {\bf B}) \times \frac{\partial {\bf B}}{\partial t}\right]   \; d^3x
\nonumber\\
&+&\int_{V} {\bf \Omega_b} \cdot
\left[ \left(\nabla \cdot \frac{\partial {\bf B}}{\partial t} \right) \; {\bf B} \right] \; d^3x
\nonumber\\
&+&\int_{V} {\bf \Omega_b} \cdot
\left[(\nabla \cdot {\bf B}) \; \frac{\partial {\bf B}}{\partial t} \right] \; d^3x
\nonumber\\
&-& \int_{V}  (\Omega_a^2 + \Omega_b^2) \; {\bf B} \cdot \frac{\partial {\bf B}}{\partial t} \;
d^3x.
\end{eqnarray}
The fourth and fifth term have the correct form. We apply the vector
identities
${\bf a} \cdot ({\bf b} \times {\bf c})={\bf b} \cdot ({\bf c} \times {\bf a})=
{\bf c} \cdot ({\bf a} \times {\bf b})$
to the first and second term
\begin{eqnarray}
\Rightarrow \frac{1}{2} \; \frac{d L}{d t} &=&
\int_{V} \left(\nabla \times \frac{\partial {\bf B}}{\partial t}\right) \cdot ({\bf B} \times  {\bf \Omega_a})  \; d^3x \nonumber\\
&+& \int_{V} \frac{\partial {\bf B}}{\partial t} \cdot
({\bf \Omega_a} \times (\nabla \times {\bf B})) \;     \; d^3x \nonumber\\
&+&\int_{V} ({\bf \Omega_b} \cdot {\bf B}) \; \nabla \cdot  \frac{\partial {\bf B}}{\partial t}
 \; d^3x \nonumber\\
&+&\int_{V} \left[{\bf \Omega_b} \;
(\nabla \cdot {\bf B})\right] \cdot \frac{\partial {\bf B}}{\partial t}  \; d^3x \nonumber\\
&-& \int_{V}  (\Omega_a^2 + \Omega_b^2) \; {\bf B} \cdot \frac{\partial {\bf B}}{\partial t} \;
d^3x.
\end{eqnarray}
Term two, four and five have the correct form. We apply
$(\nabla \times {\bf a}) \cdot {\bf b}={\bf a} \cdot (\nabla \times {\bf
b})+\nabla \cdot ({\bf a} \times {\bf b})$ to term 1 and
$\psi \nabla \cdot {\bf a}=-{\bf a} \cdot \nabla \psi+\nabla \cdot({\bf a \psi})$
to term 3
\begin{eqnarray}
\Rightarrow \frac{1}{2} \; \frac{d L}{d t} &=&
-\int_{V} \frac{\partial {\bf B}}{\partial t} \cdot [\nabla \times ({\bf \Omega_a} \times {\bf B} )]  \; d^3x \nonumber\\
&-& \int_{V} \nabla \cdot \left[({\bf \Omega_a} \times {\bf B} ) \times \frac{\partial {\bf B}}{\partial t}  \right]  \; d^3x \nonumber\\
&+& \int_{V} \frac{\partial {\bf B}}{\partial t} \cdot
({\bf \Omega_a} \times (\nabla \times {\bf B}) \;     \; d^3x \nonumber\\
&-&\int_{V} \nabla ({\bf \Omega_b} \cdot {\bf B}) \;  \cdot  \frac{\partial {\bf B}}{\partial t}
 \; d^3x \nonumber\\
&+&\int_{V} \nabla \cdot \left[ ({\bf \Omega_b} \cdot {\bf B}) \; \frac{\partial {\bf B}}{\partial
t} \right]
 \; d^3x \nonumber\\
&+&\int_{V} \left[{\bf \Omega_b} \;
(\nabla \cdot {\bf B})\right] \cdot \frac{\partial {\bf B}}{\partial t}  \; d^3x \nonumber\\
&-& \int_{V}  (\Omega_a^2 + \Omega_b^2) \; {\bf B} \cdot \frac{\partial {\bf B}}{\partial t} \;
d^3x.
\end{eqnarray}
Term one,three, four, six and seven have the correct form. We apply Gauss law
to term two and five
\begin{eqnarray}
\Rightarrow \frac{1}{2} \; \frac{d L}{d t} &=&
-\int_{V}  \frac{\partial {\bf B}}{\partial t} \cdot [\nabla \times ({\bf \Omega_a} \times {\bf B} )]  \; d^3x \nonumber\\
&-& \int_{S} {\bf \hat n} \cdot \left[({\bf \Omega_a} \times {\bf B} ) \times \frac{\partial {\bf B}}{\partial t}  \right]  \; d^2x \nonumber\\
&+& \int_{V} \frac{\partial {\bf B}}{\partial t} \cdot
({\bf \Omega_a} \times (\nabla \times {\bf B})) \;     \; d^3x \nonumber\\
&-&\int_{V} \nabla ({\bf \Omega_b} \cdot {\bf B}) \;  \cdot  \frac{\partial {\bf B}}{\partial t}
 \; d^3x \nonumber\\
&+&\int_{S} {\bf \hat n}  ({\bf \Omega_b} \cdot {\bf B}) \cdot \frac{\partial {\bf B}}{\partial
t} \; d^2x \nonumber\\
&+&\int_{V} \left[{\bf \Omega_b} \;
(\nabla \cdot {\bf B})\right] \cdot \frac{\partial {\bf B}}{\partial t}  \; d^3x \nonumber\\
&-& \int_{V}  (\Omega_a^2 + \Omega_b^2) \; {\bf B} \cdot \frac{\partial {\bf B}}{\partial t} \;
d^3x.
\end{eqnarray}
Now all terms but the second have the correct form. We apply
${\bf a} \cdot ({\bf b} \times {\bf c})={\bf c} \cdot ({\bf a} \times {\bf b})$
to the second term
\begin{eqnarray}
\Rightarrow \frac{1}{2} \; \frac{d L}{d t} &=&
-\int_{V}  \frac{\partial {\bf B}}{\partial t} \cdot [\nabla \times ({\bf \Omega_a} \times {\bf B} )]  \; d^3x \nonumber\\
&-& \int_{S} [{\bf \hat n} \times \left[({\bf \Omega_a} \times {\bf B} )\right] \cdot  \frac{\partial {\bf B}}{\partial t}    \; d^2x \nonumber\\
&+& \int_{V} \frac{\partial {\bf B}}{\partial t} \cdot
({\bf \Omega_a} \times (\nabla \times {\bf B})) \;     \; d^3x \nonumber\\
&-&\int_{V} \nabla ({\bf \Omega_b} \cdot {\bf B}) \;  \cdot  \frac{\partial {\bf B}}{\partial t}
 \; d^3x \nonumber\\
&+&\int_{S} {\bf \hat n}  ({\bf \Omega_b} \cdot {\bf B}) \cdot \frac{\partial {\bf B}}{\partial
t} \; d^2x \nonumber\\
&+&\int_{V} \left[{\bf \Omega_b} \;
(\nabla \cdot {\bf B})\right] \cdot \frac{\partial {\bf B}}{\partial t}  \; d^3x \nonumber\\
&-& \int_{V}  (\Omega_a^2 + \Omega_b^2) \; {\bf B} \cdot \frac{\partial {\bf B}}{\partial t} \;
d^3x.
\end{eqnarray}
Now all terms have the correct form and we write them more compact as
\begin{equation}
\Rightarrow \frac{1}{2} \; \frac{d L}{d t}=-\int_{V} \frac{\partial {\bf B}}{\partial t} \cdot {\bf F} \; d^3x
-\int_{S} \frac{\partial {\bf B}}{\partial t} \cdot {\bf G} \; d^2x,
\end{equation}
with
\begin{eqnarray}
{\bf F} & =& \nabla \times ({\bf \Omega_a} \times {\bf B} )
- \bf \Omega_a \times (\nabla \times \bf B)  \nonumber\\
& & +\nabla(\bf \Omega_b \cdot \bf B)-  \bf \Omega_b(\nabla \cdot \bf B)
+( \Omega_a^2 + \Omega_b^2)\; \bf B,
\end{eqnarray}
\begin{eqnarray}
{\bf G} & = & {\bf \hat n} \times ({\bf \Omega_a} \times {\bf B} )
-{\bf \hat n} (\bf \Omega_b \cdot \bf B).
\end{eqnarray}
\section{Derivation of $H$ and $I$ in (\ref{minimize2}).}
\label{appendixB}
\setcounter{equation}{0}
We vary $L$ with respect to an iteration parameter $t$ where the
magnetic field is independent from $t$ here.
\begin{eqnarray}
\frac{1}{2} \; \frac{d L}{d t} &=&
\int_{V} {\bf \Omega_a} \cdot \frac{\partial}{\partial t}
[(\nabla \times {\bf B}) \times {\bf B} -\mu_0 (\nabla P +\rho \nabla \Psi) ]  \; d^3x
\end{eqnarray}
We write the pressure $P$ and the mass density $\rho$
as functions of the number density $N$
\begin{eqnarray}
\Rightarrow \frac{1}{2} \; \frac{d L}{d t} &=&
\int_{V} {\bf \Omega_a} \cdot \frac{\partial}{\partial t}
[(\nabla \times {\bf B}) \times {\bf B}
\nonumber\\
&-&\mu_0 (k_B T \; \nabla N +
m N \; \nabla \Psi) ]  \; d^3x
\end{eqnarray}
Our aim is to write all terms as products with $\frac{\partial N}{\partial
t}$ to derive evolution equations for $N$ to minimize $L$.
The first term vanishes here because {\bf B} does not depend n $t$. We expand
the second term
\begin{eqnarray}
\Rightarrow \frac{1}{2} \; \frac{d L}{d t} &=&
-\mu_0 \;\int_{V} {\bf \Omega_a} \cdot k_B T \; \nabla \left(\frac{\partial N}{\partial t}\right)  \; d^3x \nonumber\\
&& -\mu_0 \;  \int_{V} {\bf \Omega_a} \cdot
m \; \frac{\partial N}{\partial t}\;\nabla \Psi   \; d^3x.
\end{eqnarray}
The second term has the correct form. We apply
${\bf a} \cdot \nabla \psi=\nabla \cdot({\bf a \psi})-\psi \nabla \cdot {\bf a}$
to the first term
\begin{eqnarray}
\Rightarrow \frac{1}{2} \; \frac{d L}{d t} &=&
-\mu_0 \;\int_{V} \nabla \cdot \left({\bf \Omega_a} \;  k_B T \; \frac{\partial N}{\partial t}\right) \; d^3x \nonumber\\
&& +\mu_0 \;\int_{V} \nabla \cdot {\bf \Omega_a} \; k_B T \; \frac{\partial N}{\partial t}  \; d^3x \nonumber\\
&& -\mu_0 \; \int_{V} {\bf \Omega_a} \cdot
m \; \frac{\partial N}{\partial t}\;\nabla \Psi   \; d^3x.
\end{eqnarray}
Term two and three have the correct form. We apply Gauss law to the first
term
\begin{eqnarray}
\Rightarrow \frac{1}{2} \; \frac{d L}{d t} &=&
-\mu_0 \;\int_{S} {\bf \hat n} \cdot {\bf \Omega_a} \;  k_B T \; \frac{\partial N}{\partial t} \; d^2x \nonumber\\
&& +\mu_0 \;\int_{V} \nabla \cdot {\bf \Omega_a} \; k_B T \; \frac{\partial N}{\partial t}  \; d^3x \nonumber\\
&& -\mu_0 \; \int_{V} {\bf \Omega_a} \cdot
m \; \frac{\partial N}{\partial t}\;\nabla \Psi   \; d^3x.
\end{eqnarray}
Now all terms have the correct form and we write more compact
\begin{eqnarray}
\Rightarrow \frac{1}{2} \; \frac{d L}{d t} &=&
-\int_{V} H \;  \frac{\partial N}{\partial t}   \; d^3x
-\int_{S} I \; \frac{\partial N}{\partial t} \; d^2x \nonumber\\
H &=& \mu_0 m \; {\bf \Omega_a} \cdot \nabla \Psi
-\mu_0 \; k_B T \; \nabla \cdot {\bf \Omega_a} \nonumber\\
I &=&  \mu_0 \; k_B T \; {\bf \Omega_a} \cdot {\bf \hat n}.
\end{eqnarray}

\section{Derivation of {\bf H} and {\bf I} in (\ref{minimize3}).}
\label{appendixC}
\setcounter{equation}{0}
An essential advantage in the derivation of the variational derivative
of $G$ in (\ref{tomoeq}) is the fact that $G$ is convex quadratic expression.
While $I^{\rm obs}$ is just a data vector,
${\cal I}$ is a linear operator from the model space $\{N(V)\}$ into
the data space $\{I_{p,i}\}$:
\begin{equation}
 {\cal I}_{p,i}(N)=\int\limits_{{\cal C}_{p,i}} N(\mathbf{x})\,d\ell\quad:\quad
 \mbox{model space}\longrightarrow\mbox{data space}.
\label{forwop}\end{equation}
The beam was defined in the main text as ${\cal C}_{p,i}$ =
$\{\mathbf{x}\in V\,|\,\mathbf{x}$ projects onto pixel $p$ for the
view direction of image $i$ $\}$.
The adjoint of (\ref{forwop}) is
\begin{equation}
 {\cal J}(\mathbf{x}, I)=\sum\limits_{p,i}
 \delta_{{\cal C}_{p,i}}(\mathbf{x}) \, I_{p,i}\quad:\quad
 \mbox{data space}\longrightarrow\mbox{model space}
\label{adjntop},
\end{equation}
where
$\delta_{{\cal C}_{p,i}}(\mathbf{x})$ = 1 for $\mathbf{x}\in{\cal C}_{p,i}$ and
$\delta_{{\cal C}_{p,i}}$ = 0 else.
The adjointness can easily be checked by insertion of the respective
definitions into
\begin{equation}
\sum\limits_{p,i} I_{p,i} \, {\cal I}_{p,i}(N) =
\int\limits_V {\cal J}(\mathbf{x}, I) \, N(\mathbf{x}) \,d^3x
\end{equation}
Now minimizing (\ref{tomoeq})
\begin{equation}
G(N)=\sum\limits_{p,i}
  |I_{p,i}^{\rm obs}-{\cal I}_{p,i}(N)|^2
  +\mu \int\limits_V R(N)^2\, d^3x
\end{equation}
with (\ref{bfldreg})
\begin{equation}
 R(N) = \frac{1}{k_B T} (\hat{\mathbf{B}}\cdot\nabla) P
     + \rho(\hat{\mathbf{B}}\cdot\nabla) \Psi
      = (\hat{\mathbf{B}}\cdot\nabla) N
      - \frac{\hat{\mathbf{B}}\cdot\hat{\mathbf{r}}}{H_0} N
\end{equation}
yields
\begin{eqnarray*}
\Derv{G}{t}&=&2\sum\limits_{p,i}
  ({\cal I}_{p,i}(N)-I_{p,i}^{\rm obs}) {\cal I}_{p,i}(\Dpar{N}{t})
\\
 &+&2\mu \int\limits_V R\;\left[
        (\hat{\mathbf{B}}\cdot\nabla) \Dpar{N}{t}
      - \frac{\hat{\mathbf{B}}\cdot\hat{\mathbf{r}}}{H_0} \Dpar{N}{t}
                   \right] \,d^3x
\\
&=&2\int\limits_V
  {\cal J}(\mathbf{x}, {\cal I}_{p,i}(N)-I_{p,i}^{\rm obs})
  \,\Dpar{N}{t}
\\
 &-&2\mu \int\limits_V \left[(\hat{\mathbf{B}}\cdot\nabla)R
 + R\frac{\hat{\mathbf{B}}\cdot\hat{\mathbf{r}}}{H_0}\right]
         \Dpar{N}{t} \,d^3x
\\
 &+&2\mu \int\limits_S R\;(\hat{\mathbf{B}}\cdot\hat{\mathbf{n}}) \Dpar{N}{t}\,d^2x
\\
&=&\int\limits_V H \Dpar{N}{t} \; d^3x
  +\int\limits_S I \Dpar{N}{t} \; d^2x,
\end{eqnarray*}
where $\hat{\mathbf{n}}$ is the unit normal on the surface $S$ and
\begin{eqnarray}
H(\mathbf{x}) &=& 2\sum\limits_{p,i} \delta_{{\cal C}_{p,i}}(\mathbf{x})
                ({\cal I}_{p,i}(N)-I_{p,i}^{\rm obs})
\nonumber \\
 &-&2\mu \; \big[(\hat{\mathbf{B}}\cdot\nabla)R(N)
 + \frac{\hat{\mathbf{B}}\cdot\hat{\mathbf{r}}}{H_0} R(N)\big],
\\
I(\mathbf{x}) &=&
 2\mu \; (\hat{\mathbf{B}}\cdot\hat{\mathbf{n}}) R(N).
\end{eqnarray}

\end{article}
\end{document}